# A new apparatus for deep patterning of beam sensitive targets by means of high-energy ion beam

A. Rovelli,[a] A. Amato,[a] D. Botta,[b] A. Chiodoni,[b] R. Gerbaldo,[b] G. Ghigo,[b]
L. Gozzelino,[b] F. Laviano,[b] B. Minetti,[b] and E. Mezzetti[b]

*[a]INFN - Laboratori Nazionali del Sud, Via S. Sofia 44, 95100 Catania, Italy*

*[b]Dept. of Physics, Politecnico di Torino and INFN - Sez. di Torino, C.so Duca degli Abruzzi, 24, 10129 Torino, Italy*

## Abstract

The paper reports on a high precision equipment designed to modify over 3-dimensions (3D) by means of high-energy gold ions the local properties of thin and thick films. A target-moving system aimed at creating patterns across the volume is driven by an x-y writing protocol that allows one to modify beam sensitive samples over micrometer-size regions of whatever shape. The apparatus has a mechanical resolution of 15 nm. The issue of the local fluence measurement has been particularly addressed. The setup has been checked by means of different geometries patterned on beam sensitive sheets as well as on superconducting materials. In the last case the 3D modification consists of amorphous nanostructures. The nanostructures create zones with different dissipative properties with respect to the virgin regions. The main analysis method consists of magneto-optical imaging that provides local information on the electrodynamics of the modified zones. Features typical of non-linear current flow hint at which pattern geometry is more functional to applications in the framework of nanostructures across superconducting films.

## 1. Introduction

The high rate of development of beam particle accelerators in relation to the issue of modifying material properties by means of "writing" technologies (based on ion-induced defects), claims the extension of the related techniques to high-energy ion beams. Swift ion beams enable one to modify structural and electrodynamics properties up to a deeper level then keV accelerators [1]. The modifying effects involve indeed the three-dimensional structure of a large class of thin and thick films [2, 3, 4, 5, 6, 7, 8, 9].

The most updated challenge consists in confining into microsize regions the ion-affected zones by means of tailored irradiation protocols [10]. From a macroscopic electrodynamics point of view, this issue is aimed at either simulating damage or at modulating film properties over micrometric scale. The property changes are permanently "written" into the target [11]. In particular, in superconductor films, the 3D modification consists of a distribution of amorphous nanostructures [12]. The nanostructures set up zones with different dissipative properties with respect to the virgin regions: different field penetration and Meissner current distribution are created. In such a way an external perturbation will be localized through the change of a dissipative signal across the modified zone.

In this paper we report on an apparatus to modulate the properties of beam sensitive samples by means of 4.2 GeV gold ions. The apparatus was built at INFN–LNS (Laboratori Nazionali del Sud), Catania, Italy [13].

The experimental set up is described in Sec. 2. This section reports on the solutions chosen in order to guide the beam towards a stainless steel disk with micrometric pinhole aperture, to evaluate the local ion fluences and, finally, to drive the x-y step motion of the apparatus. Sec. 3 provides more details on the issues of beam calibration and local fluence evaluation as well as on readouts and feedback controls. In Sec. 4 the apparatus reliability to generate micropatterns of any programmed shape is demonstrated. In particular Sec. 4.1 concerns results obtained over polyester sheets, to simulate the geometry of local damage over any kind of beam sensitive material, as for instance semiconducting chips or magnetic materials. In Sec 4.2 we focus on high temperature superconducting (HTSC) films in order to observe shapes resulting by a given modifying geometry. Finally in Sec 5 concluding remarks are reported.

## 2. The experimental set-up.

### 2.1 Main components.

A dedicated in-vacuum irradiation apparatus for high-energy heavy-ion patterning has been designed to be connected to the downstream part of the accelerator user lines. Figure 1 shows an image and the corresponding schematic drawing of the experimental set-up of this apparatus. Basic components are two cross-mounted MICOS Linear Stages LS-110 controlling the bi-directional movement of the target in a plane perpendicular to the ion beam direction. Each stage has an optical encoder that, once connected to the motor controller electronics Micos SMC Taurus, enables the target to move with a resolution of 15 nm over a total length of 28 mm. The target and a scintillating screen ($Al_2O_3$, Morgan Matroc Chromox-6, 0.7 mm thick) are positioned on a 9 mm thick lead-glass sample holder fixed to the moving system. A control CCD camera (B/W CCD camera WAT-902H CCIR equipped with a lens of 16 mm and an extension tube of 20 mm exhibiting a viewing area of 6.5 x 4 $mm^2$) is positioned close to the backside of the sample holder and is centered with respect to the ion beam direction. This CCD camera can visualize the ion beam spot through the scintillating screen. It is connected with a frame grabber NI-PCI 1411 installed into the control PC. Two Faraday Cups allow one to monitor and measure the fluences.

### 2.2 The fluence measurement system.

The fluence measurement system consists of a suitably assembled Faraday Cup (henceforth labeled Faraday Cup #1) and the related read-out electronics. A standard Faraday Cup (henceforth labeled Faraday Cup #2) positioned upstream the apparatus is used for shutting down the beam during the irradiation as well as for providing reference beam current evaluation. The Faraday Cup #1 has been designed accordingly to the irradiation protocol requirements: it consists of a Faraday Cup with a threaded hole (Ø 15 mm) in the bottom where a beam collimator with changeable stainless steel disks with pinhole aperture can be screwed. The pinhole diameter, setting the beam spot size, is chosen on the basis of the irradiated pattern requirements. This Faraday Cup is positioned in front of the sample holder and is aligned with respect the beam direction (Fig. 1). During the beam characterization, before the sample irradiation, the threaded hole conveys the beam towards a scintillating screen fixed on the sample holder. When the stainless steel disk with the pinhole is fixed, a beam spot of known shape and size will be set for drawing the irradiation pattern.

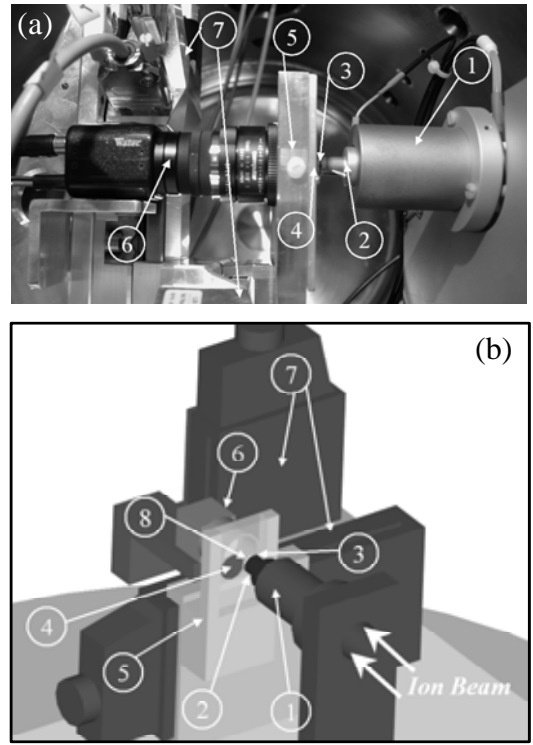

Fig. 1 – (a) Image and (b) corresponding schematic drawing of the experimental arrangement of the irradiation set up for high-energy heavy-ion patterning: (1) Faraday Cup #1 – (2) Collimator – (3) Stainless steel disk with micrometric pinhole aperture – (4) Scintillating screen – (5) Lead-glass sample holder – (6) CCD camera – (7) Linear stages – (8) Sample.

Both the Faraday Cups are connected to an electronic board designed for this application (Fig. 2(a)). It is equipped with two input channels: the first one is a current/voltage converter (gain factor = 1 nA/ 1 V) used for reading the reference Faraday Cup #2, the second one is a charge/pulse converter (1 pC/pulse) used for integrating the beam current during the irradiation. The typical beam current for the irradiation is lower than 1 nA. The charge/pulse converter has been checked for current values ranging from 1 pA to 10 nA. The linearity error is 0.002 % FSR.

The electronic board is connected with a NI-PCI 6024E DAQ board installed into the local PC that also drives the motor controllers. This DAQ board is equipped with two 20 MHz 24-bit up/down counter/timers used for reading the integrated charge; it also has one 12-bit ADC and three 8-bit digital I/O ports employed in reading the ion beam current signal coming from the reference Faraday Cup #2 and in controlling the beam shutter, respectively.

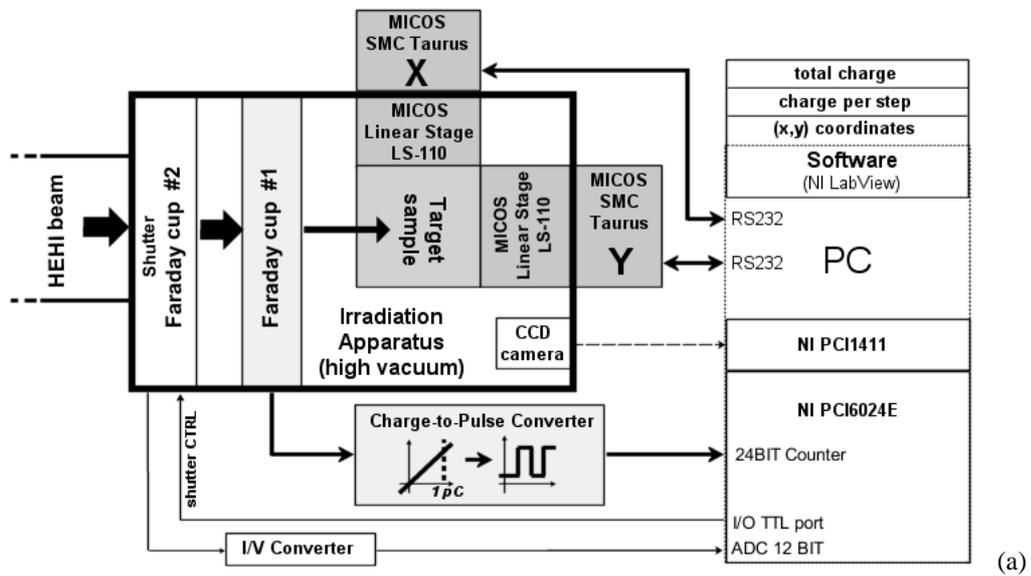

(a)

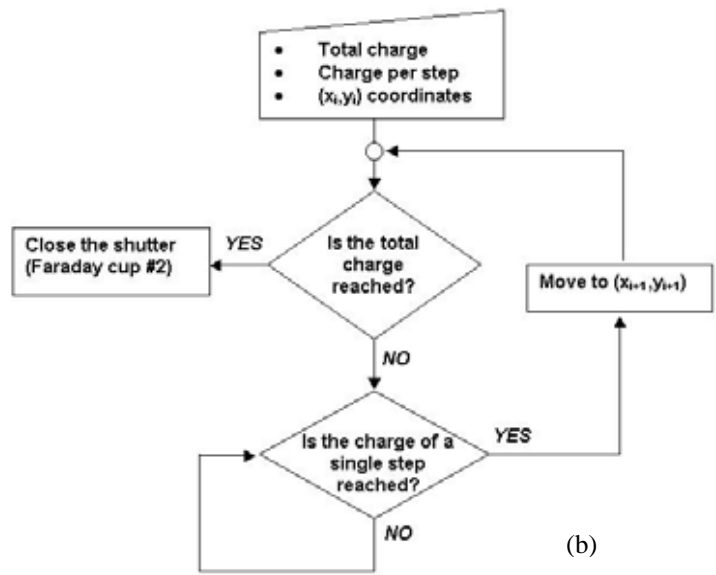

(b)

Fig. 2 – (a) Block diagram of the main components of the irradiation apparatus. (b) Flow chart of the software for the read-out and feedback controls.

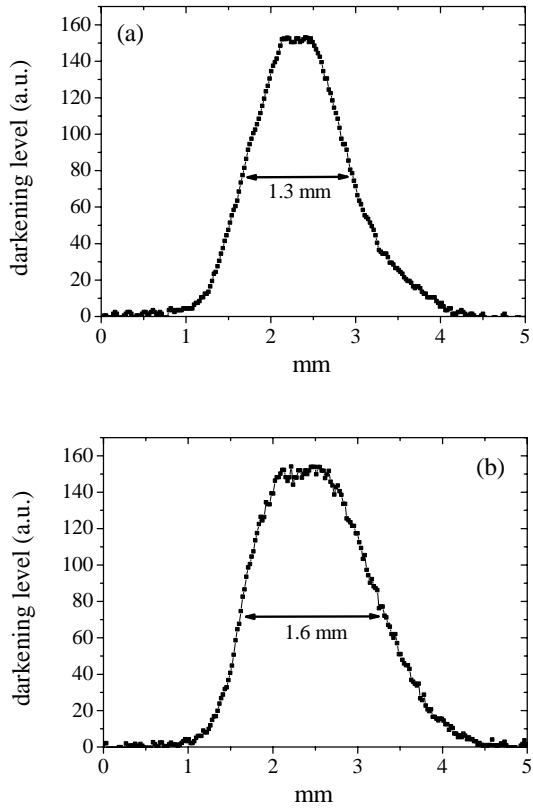

Fig. 3 - Beam spot profiles evaluated from the darkening level of an ion-beam sensitive polyester sheet (see text) along the (a) horizontal and (b) vertical directions of movement.

## 3. The irradiation protocol.

### 3.1 Beam calibration and fluence evaluation.

The first step of the beam calibration is the beam tuning aimed at achieving a suitable homogeneity. During this operation the ion beam, passing through the Faraday Cup #1 and the threaded hole in the bottom side, hits the scintillating screen positioned onto the lead-glass sample holder. The CCD camera visualizes the transmitted light. The output signal of the CCD camera is analyzed on the control PC and is used for a first optimization of the beam conditioning process. Then the beam shape and the ion fluence distribution are visualized by irradiating an ion-beam sensitive polyester sheet. The darkening level is proportional to the local irradiation fluence in the investigated range of fluence (Fig. 3).

Once optimal beam conditions are steadily obtained, the stainless steel disk with the pinhole is screwed on the bottom side of the Faraday Cup #1. In this configuration the very small part of the beam passing through the pinhole aperture hits the scintillating screen and is visualized by the CCD camera. The user, by means of a digital zoom, can verify whether this micrometric beam spot is centered with respect to the previously measured millimetric beam spot. The (x,y) coordinates and the shape of the beam spot are then stored for the on-line cross check.

After this beam calibration it is possible to start the irradiation procedure, consisting into combining the fluence evaluation with the x-y writing mode.

The irradiation fluence, $\phi$, on the target is proportional to the charge, $q_{hole}$, released onto it:

$$\phi = q_{hole}/(n \cdot e \cdot A) \qquad \text{(eq. 1)}$$

where $n$ is the ion charge, $e$ is the elementary charge and $A$ is the pinhole aperture area.

The apparatus allows one to measure the ion current inside the Faraday Cup #1 and, by integration, the charge, $q_{FC1}$, released onto it. From the two-dimensional beam spot profiles shown in Fig. 3 we calculate the two dimensional charge probability density function $f(x, y)$. Thus the relation between $q_{FC1}$ and $q_{hole}$ results to be:

$$q_{FC1} = q_{hole}\left(1 - \int\int_{hole} f(x, y)dxdy\right) \qquad \text{(eq. 2)}$$

It turns out that $q_{hole}$ is about 0.2 % of $q_{FC1}$ for a 70 μm pinhole aperture and becomes about 0.04 % for a 30 μm pinhole aperture.

In the x-y writing mode the irradiation patterns are obtained by a sequence of linear steps: each x-y step is a fraction of the pinhole aperture diameter. This fraction is suitably chosen to draw patterns as smooth as possible. Two kinds of irradiations are possible: irradiation at constant dose and irradiation with a given dose profile. In the first case the charge per single step is constant, while in the second case the charge per single step follows a tailored law [14].

The velocity of the patterning process depends on the chosen ion beam current as well as on the irradiation fluence. The time of a single step movement is indeed negligible (typically $10^3$-$10^4$ μm/s). Typical values of current and fluence are 650 pA and $10^{11}$ ions/cm², respectively. With these values, when the beam is collimated through a circular 30 μm wide pinhole, the patterning-process velocity is 1 μm/s.

Fig. 2(b) shows the flow chart of the control software. The movement of the two linear MICOS stages is controlled in feedback with the read-out signal of the Faraday Cup #1, by means of a NI Labview-based control software. This home-made software requires, as input parameters, the value of the total charge, a matrix containing the x-y coordinates of each step and the charge per step. As above mentioned the ion beam current output signal of the Faraday Cup #1 is converted into a frequency signal and then in TTL pulses. When the number of generated TTL pulses (i.e. pC of charge) reaches the charge per step previously set by the user, the software gives the feedback signal (through two RS-232 ports) to move the linear stages on the next set of (x,y) coordinates. A new value of the charge per step can be set, if needed, until the total charge is reached.

## 4. Experimental results

### 4.1 Beam-sensitive sheets

The x-y writing equipment has been tested by means of 4.2 GeV Au-ions at the Superconducting Cyclotron (CS) facility of INFN-LNS. In the first place the targets were the already mentioned polyester sheets. We patterned several x-y shapes with different geometries (Fig. 4). The map of the ion fluence distributions on these irradiated sheets is reconstructed under microscope on the basis of the local value of the darkening level, see Figs. 4 (a)-(d). A representative irradiation fluence profile across a side of a square perimeter is reported in Fig. 4(e). Due to the collimator shape (pinhole), the irradiation fluence profile shows a bell-shape with a small plateau in the central part where the fluence is maximum. It turns out that local damage can be simulated over beam-sensitive materials in micrometer-size zones, with an accuracy evaluated below. (Sec 4.2). For any kind of experiment the irradiation of these sheets is routinely performed before any change of pattern geometry.

### 4.2. Superconducting materials

YBa$_2$Cu$_3$O$_{7-x}$ (YBCO) superconducting films were grown by means of pulsed laser deposition technique on yttria stabilized zirconia (YSZ) substrate, with a 10 nm thick CeO$_2$ buffer layer. The critical current density is about $3 \cdot 10^{11}$ A/m$^2$ at T = 5 K. All the films are 0.3 μm thick. In this paper we consider two (9x9) mm$^2$ YBCO samples labeled YBCO#1 and YBCO#2, respectively.

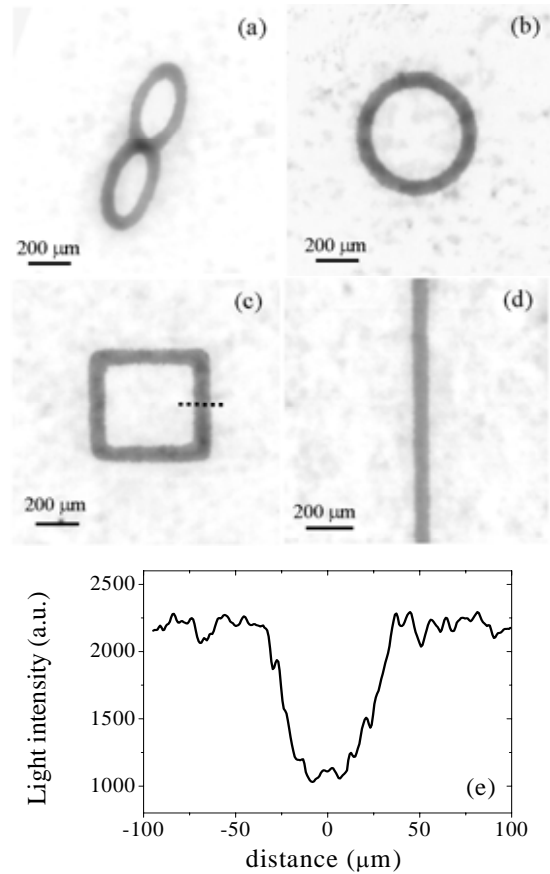

Fig. 4 - (a)-(d) Optical image of polyester sheets micro-patterned by means of 4.2 GeV Au-ions following different geometries; (e) beam profile distribution along the dotted line of Fig. 4(c). These patterns were obtained by collimating the beam through a circular 70 μm wide pinhole. The spikes in the curves are due to the granularity of the polyester sheet surface.

The two YBCO films were irradiated with 4.2 GeV Au-ions. The beam was always directed perpendicular to the film surface. 4.2 GeV Au-ions cross the YBCO film and the CeO$_2$ buffer layer and implant in the YSZ substrate at a depth of 117.0 μm, corresponding to about 23% of the substrate thickness (Monte-Carlo TRIM code simulation [15]). The energy, E$_p$, released by the ions is almost constant along the whole YBCO film thickness and is high enough - E$_p$ ~ 35 MeV/(ion μm) - to induce columnar tracks [16]. The damage caused by the ion collision is visible in Fig. 5 where a planar view of the YBCO#1 sample surface, obtained by transmission electron microscopy (TEM) analysis, is reported. These extrinsic defects are insulating columns with a diameter of a few

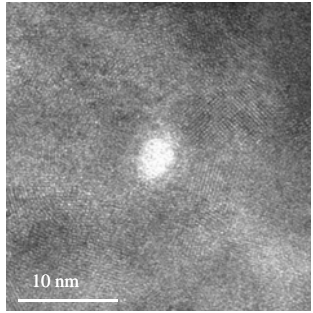

Fig. 5 – High-resolution TEM image of a [001] planar view of the YBCO#1 sample surface with a columnar defect (white area) produced by the Au ion irradiation. The column is surrounded by an amorphous region, which results in enlarged and inhomogeneous defected zone.

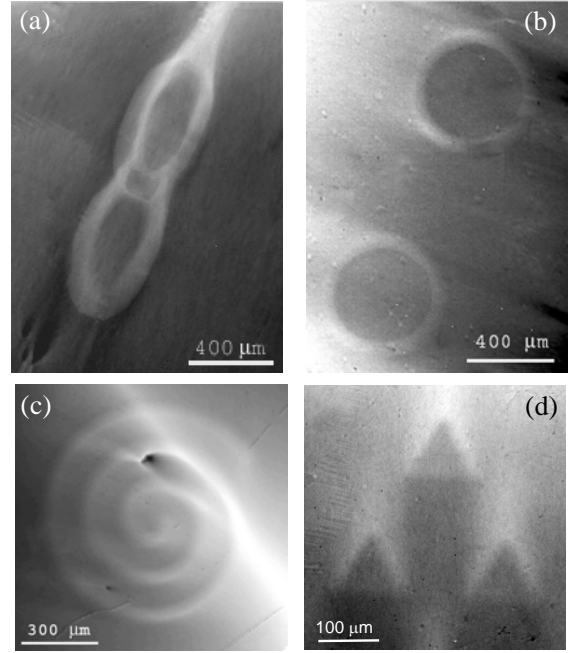

Fig. 6 - Magneto-optical images of sample YBCO#2 after ion patterning with different geometries (a) "infinity", (b) rings, (c) spiral and (d) triangles. These patterns were obtained by means of 4.2 GeV Au-ions ($\phi = 1.0 \cdot 10^{11}$ ion/cm$^2$ corresponding to $B_\phi = 2$ T [25]) by collimating the ion beam through a circular 70 μm large pinhole (Figs. (a), (b), (c)) and through a circular 30 μm large pinhole (Fig. (d)). The images were taken at T = 50 K in remnant state after a field cooling in a magnetic field of 80 mT. Brighter colours: higher magnetic flux density regions (higher remnant magnetic field), darker colours: lower magnetic flux density regions.

nanometers.

To visualize into superconducting films the patterned regions whose signature is the induced modulation of the magnetic field, magneto-optical analysis was performed on the YBCO#2 sample [17, 18, 19]. The magneto-optical signal is obtained by coupling the superconductor with a ferromagnetic film whose domain orientation is decorated by the superconducting electro-dynamical signal. The change of the domain orientation is read as a rotation of a monochromatic light polarization plane [20].

In Fig. 6 magneto-optical images of different pattern geometries ("infinity", rings, spiral, triangles), observed by magneto-optical analysis, are reported. The resolution of the MO imaging with the used optical configuration is 1 μm. Within this resolution we did not find any misplace of the irradiation patterns due to the automatic patterning set up, so we speculate that the accuracy of our writing system is either equal or even better than our MO resolution, it means at least 1 μm. Inside the superconducting condensate, the mentioned patterns behave as macroscopic defects [21]. Many features typical of a non linear current flow (very similar to the charge flow in confined plasma [21]) can be observed by MO technique, which indeed reveals long range disturbances of the normal component of **B**(x,y). Typically with respect to the considered shape, characteristics magnetic flux jets are observed as flame-like patterns. These effects must be suitably controlled by changing the geometrical constrains [22].

## 5. Discussion and conclusions.

High-energy ions modify the 3D structural properties of film systems. The confinement of the ion-affected region at micrometric scale allows one to address different issues: for example to simulate local damage on materials, to modify magnetic material properties [23, 24] and finally, in superconductors, to confine external excitations (for instance field-, particle- or photon- induced). Looking at these applications, we developed a novel apparatus to confine the high-energy heavy-ion beam into a micrometric size spot (down to 30 μm up to now) and to control with nanometric resolution the two-dimensional movement of the target with respect to the collimated heavy-ion beam.

The facility enables one to automatically perform x-y writing over 28x28 mm$^2$ regions with a 15 nm resolution. Variable dose profiles characterized by tailored gradients are easily performed with the same resolution level. The drawn patterns also include non-simply connected geometries, not obtainable by means of static irradiation through laser-cut masks. Moreover the apparatus allows to easily program any kind of checkboard-like patterns. Scaling down of the macrodefect characteristic size is currently in progress.

To simulate localized damage in any kind of ion-beam-sensible material, beam sensible sheets were chosen and patterns of micrometer size were shown.

Finally the issue of localized patterning of superconductors has been addressed. Local analysis were performed by magneto-optical technique. The not simply-connected patterns exhibit magnetic flux jets that are dependent on the geometric constraints.

## Acknowledgements

We acknowledge the contribution of the Istituto Nazionale di Fisica Nucleare (INFN) under the La.Sca.R. (Large Scale Radiation Damage) and Na.St.R.I. (Nanostructure Radiation Induced) projects. The support of the INFN-LNS staff during the irradiation runs is gratefully acknowledged. We thank W. K. Kwok for helpful discussions. The European Science Foundation (E.S.F.) Vortex Programme is also gratefully acknowledged.